\documentclass[preprint,12pt]{elsarticle}

\usepackage{amssymb}
\usepackage{amsmath,amsfonts}
\usepackage{algorithmic}
\usepackage{array}
\usepackage[caption=false,font=normalsize,labelfont=sf,textfont=sf]{subfig}
\usepackage{textcomp}
\usepackage{stfloats}
\usepackage{url}
\usepackage{verbatim}
\usepackage{graphicx}
\def\BibTeX{{\rm B\kern-.05em{\sc i\kern-.025em b}\kern-.08em
    T\kern-.1667em\lower.7ex\hbox{E}\kern-.125emX}}
\usepackage{balance}

\usepackage{booktabs}                        
\usepackage{subcaption}   
\usepackage{color}
\usepackage{amsmath}
\usepackage{amssymb}
\usepackage{graphicx}
 \usepackage{dsfont}
\usepackage{amsfonts}
\usepackage{amsthm} 
\usepackage{stmaryrd}
\usepackage[all]{xy}
\usepackage{multirow}

\usepackage{qcircuit}
 
\def\>{\ensuremath{\rangle}}
\def\<{\ensuremath{\langle}}

\newcommand{\hs}{\mathcal{H}}

\newtheorem{thm}{Theorem}[section]

\newtheorem{lem}{Lemma}[section]
\newtheorem{defn}{Definition}[section]

\newtheorem{exam}{Example}[section]

\journal{}

\begin{document}

\begin{frontmatter}

\title{Reasoning about Shared-Qubit Concurrency in Quantum Computing}

\author{Mingsheng Ying}

\affiliation{organization={Centre for Quantum Software and Information, University of Technology Sydney}, 
           addressline={15 Broadway}, 
           city={Ultimo},
           postcode={2007}, 
          state={NSW},
            country={Australia}}

\begin{abstract}
In this paper, we develop a semantic model for the parallel execution of quantum circuits, providing a formal foundation for studying shared-qubit concurrency in quantum computing. As in classical shared-memory concurrency, parallel quantum circuits may interfere with one another through operations on shared qubits. However, quantum entanglement and the noncommutativity of quantum operations introduce additional challenges for reasoning about their correctness. To address these challenges, we generalise the classical Owicki–Gries method to quantum computing at the level of quantum circuits. Our approach enables compositional reasoning about parallel quantum circuits by separating the verification of individual components from the analysis of interference between them. We formulate proof rules for establishing the correctness of parallel quantum circuits and prove their soundness with respect to the proposed semantic model.
\end{abstract}


\end{frontmatter}

\section{Introduction}

Shared-memory concurrency is an important technique in classical programming, allowing multiple components of a program to cooperate through a common memory. It provides a natural way to express computations in which different tasks access and modify shared data. However, designing and reasoning about shared-memory parallel programs are notoriously challenging. An internal survey at Microsoft \cite{Microsoft} revealed that concurrency is widely used within the company. However, 60\% of respondents reported frequently encountering concurrency-related issues, and most concurrency bugs were considered to be of high severity.
 
In a shared-memory parallel program, an action performed by one component may modify the shared state on which another component depends, thereby invalidating assumptions that would hold during its isolated execution. Consequently, the behaviour of a parallel program depends not only on the computations performed by its individual components, but also on the order in which their actions are interleaved.

A similar challenge arises in quantum computing when the components of a parallel quantum circuit operate on shared qubits. For components acting on disjoint registers, their operations commute, and their combined behaviour can be described by the tensor product of their respective transformations. When registers overlap, however, an operation performed by one component may alter the quantum state subsequently accessed by another. Different execution orders can therefore produce different transformations and different output states. To see this more clearly, let us consider a simple example: 

\begin{exam}\label{exam-1} Let $\mbox{SWAP}$ and $H$ stand for the swapping gate and Hadamard gate, respectively. Consider the following parallel execution of two quantum circuits performed on a system of three qubits $q_1,q_2,q_3$:
$$P\equiv \mbox{SWAP}[q_1,q_2]\ \|\ H[q_1].$$ Note that quantum gates $\mbox{SWAP}$ and $H$ in $P$ share qubit $q_1$. 
The actual implementation of $\mbox{SAWP}$ on hardware may be:
$$\mbox{SWAP}^\prime[q_1,q_2]\equiv\mbox{CNOT}[q_1,q_2];\mbox{CNOT}[q_2,q_1];\mbox{CNOT}[q_1,q_2].$$ Let us write: 
$$P^\prime\equiv \mbox{SWAP}^\prime[q_1,q_2]\ \|\ H[q_1].$$ Then the question is whether $P$ and $P^\prime$ are equivalent? The answer is No. Indeed, if $\mbox{SWAP}$ is considered as an atomic action, 
then executing $P$ with input $|1\rangle_{q_1}|0\rangle_{q_2}|0\rangle_{q_3}$ yields final state $|+\rangle_{q_1}|1\rangle_{q_2}|0\rangle_{q_3}$ or $|0\rangle_{q_1}|-\rangle_{q_2}|0\rangle_{q_3}$, where $|\pm\rangle=\frac{1}{\sqrt{2}}(|0\rangle\pm |1\rangle)$. 
However, the following execution of $P^\prime$ with the same input $|1\rangle_{q_1}|0\rangle_{q_2}|0\rangle_{q_3}$ yields a final state that is different from any final state of $P$: 
\begin{equation*}\label{f-state}|1,0\rangle|0\rangle\stackrel{\mbox{CNOT}[q_1,q_2]}{\rightarrow}|1,1\rangle|0\rangle\stackrel{H[q_1]}{\rightarrow}|-,1\rangle|0\rangle\stackrel{\mbox{CNOT}[q_2,q_1]}{\rightarrow} -|-,1\rangle|0\rangle\stackrel{\mbox{CNOT}[q_1,q_2]}{\rightarrow}\frac{1}{\sqrt{2}}(|10\rangle-|01\rangle)|0\rangle.\end{equation*} It is interesting to note that the final state of $P$ are all product state, but the final state of $P^\prime$ in (\ref{f-state}) is an entangled state.
\end{exam}

From the above example, we observe that, as in classical computing, the individual components of a parallel quantum circuit can interfere with one another by updating shared qubits. Thus, the input/output behaviour of a parallel quantum circuit is not determined solely by the input/out behaviour of its components. On the other hand, quantum memory introduces features that distinguish this problem from its classical counterpart. In particular, shared registers may be entangled with other registers, so reasoning about the reduced state of a shared register alone can lose correlations relevant to the computation. An operation on one register can change these correlations even when the reduced state of another register remains unchanged.

Shared-qubit concurrency thus raises a fundamental question: \begin{itemize}\item\textit{How can we understand and reason about the behaviour of a quantum computation whose components interact through a common quantum memory?}
\end{itemize}

In classical computing, one of the major approaches to reasoning about shared-memory parallel programs is the Owicki–Gries method \cite{OG76a, OG76} (see also Chapters 7 and 10 of the textbooks \cite{Apt, Cverify}, respectively) . Its central idea is to separate the verification of the correctness of individual components from checking that they do not interfere with one another. First, each component is annotated with assertions and verified using Hoare logic as if it were executed in isolation. Then, interference freedom is established by checking that every atomic action of one component preserves the assertions used in the proofs of the other components, whenever both the action’s precondition and the relevant assertion hold. These checks ensure that each component’s correctness proof remains valid under arbitrary interleavings of the components’ atomic actions. The method therefore combines local correctness proofs with explicit compatibility conditions to establish the correctness of the parallel program as a whole. Formally, the Owicki–Gries method can be presented as the proof rule (R-PC) in Figure \ref{fig -1}, 
\begin{figure}[h]\centering
\begin{align*}(\mbox{R-PC})\ \ \ \frac{{\rm Proofs\ of}\ \left\{\varphi_i\right\}P_i\left\{\psi_i\right\}\ (i=1,...,n)\ {\rm are\ interference\ free}}{\left\{\bigwedge_{i=1}^n \varphi_i\right\}P_1\|\cdots\|P_n\left\{\bigwedge_{i=1}^n \psi_i\right\}}
\end{align*}
\caption{Proof Rule for Parallel Composition.}\label{fig -1}
\end{figure}
where $P_1\|\cdots\|P_n$ is a shared-memory parallel program. In the premises of this rule, the correctness of each component $P_i$ is expressed by a Hoare triple $\{\varphi_i\}P_i\{\psi_i\}$, where the precondition $\varphi_i$ and postcondition $\psi_i$ are first-order logical formulas. Subject to the noninterference conditions, the correctness of the parallel program $P_1\|\cdots\|P_n$ is then inferred, with precondition $\bigwedge_{i=1}^n\varphi_i$ and postcondition $\bigwedge_{i=1}^n\psi_i$.

\paragraph{\textbf{Contributions of this paper:}} The aim of this paper is to generalise the Owicki–Gries method to quantum computing. We achieve this goal for the special case where $P_1,\ldots,P_n$ are quantum circuits without mid-circuit measurements. In this setting, following the approach of Birkhoff–von Neumann quantum logic \cite{BvN36}, it suffices to use subspaces of the underlying Hilbert spaces as preconditions and postconditions, since we are concerned with characterising the possible output pure quantum states arising from a given set of input pure quantum states. The main technical challenge is to develop a satisfactory treatment of interference that accounts for both the ordering of operations and their effects on the joint quantum state, which may be entangled.

However, we emphasise that extending this approach to general quantum programs with control flow, such as conditional statements based on mid-circuit measurements and loops, remains a largely open problem. In this more general setting, correctness is concerned not only with the possible output states but also with the expected values of observables in the output states, as specified by quantum Hoare triples whose preconditions and postconditions are general Hermitian operators \cite{DP06, Ying11}.

\paragraph{\textbf{Organisation of the paper:}} The remainder of this paper is organised as follows. Section \ref{sec-prel} reviews some technical preliminaries. Section \ref{sec-qc} defines the syntax and semantics of quantum circuits and introduces a semantic model for their parallel execution. Section \ref{sec-correct} defines correctness formulas for quantum circuits. Section \ref{sec-proof-qc1} presents proof rules for reasoning about the sequential execution of quantum circuits. Section \ref{sec-atom} develops a notion of proofs for quantum circuits containing atomic regions. The main result of this paper is presented in Section \ref{sec-proof-qc2}, where we develop proof rules for reasoning about the parallel execution of quantum circuits with shared qubits. Section \ref{sec-case} illustrates the application of these rules through a case study. Section \ref{sec-related} discusses related work, and Section \ref{sec-con} concludes the paper.

\section{Preliminaries}\label{sec-prel}

In this section, we briefly review the basic ideas of Birkhoff–von Neumann quantum logic \cite{BvN36} as technical preliminaries. This logic will serve as the assertion language for expressing the preconditions and postconditions of quantum Hoare triples used to specify the correctness of the quantum circuits studied in this paper. We then establish several technical lemmas needed to prove the soundness of our proof rules.

Since we only consider quantum circuits in this paper, all Hilbert spaces are assumed to be finite-dimensional.

\subsection{Birkhoff-von Neumann Quantum Logic}
Recall from classical logic that a proposition about a classical system can be mathematically modelled by a subset $X\subseteq\Omega$, where $\Omega$ is the state space -- the set of all possible states -- of the system. For each $\omega\in\Omega$, satisfaction $\omega\models X$ (state $\omega$ satisfies proposition $X$) is interpreted as $\omega\in X$. In quantum logic, however, all proposition about a quantum system are mathematically represented by a subspace of the state space of the system, which is a (finite-dimensional) Hilbert space $\hs$. For any pure quantum state $|\psi\rangle\in\hs$ and subspace $X$ of $\hs$, satisfaction $|\psi\rangle\models X$ is also interpreted as membership $|\psi\rangle\in X$.    

For classical propositions, the logical connectives \textquotedblleft and\textquotedblright, \textquotedblleft or\textquotedblright, \textquotedblleft not\textquotedblright\ can be simply interpreted as set-theoretic operations $\cap,\cup$ and $^c$ (complement), respectively. However, this interpretation of connectives cannot be directly generalised to the quantum case because although $\cap$ preserves closeness under linear combination, $\cup$ and $^c$ do not.  
The operations appropriate for the interpretations of logical connectives \textquotedblleft and\textquotedblright, \textquotedblleft or\textquotedblright, \textquotedblleft not\textquotedblright\ for quantum propositions are $\cap, \vee$ (join) and $^\perp$ (ortho-complement) \cite{BvN36}, where for any subspaces $X,Y$ of $\hs$, 
\begin{itemize} 
\item $X\vee Y=\mathrm{span} (X\cup Y)$, where for any subset $Z\subseteq\hs$, $\mathrm{span}\ Z$ is the smallest subspace of $\hs$ containing $Z$:
\begin{align*}\mathrm{span}\ Z&=\left\{\sum_{i=1}^n\alpha_i|\psi_i\rangle:n\geq 1, \alpha_i\in\mathbb{C}\ {\rm and}\ |\psi_i\rangle\in Z\ {\rm for\ all}\ 1\leq i\leq n \right\}.\end{align*}
\item $X^\perp =\{|\psi\rangle\in\hs:|\psi\rangle\bot|\varphi\rangle\ {\rm for\ all}\ |\varphi\rangle\in X\},$ where $|\psi\rangle\bot |\varphi\rangle$ means that $|\psi\rangle$ and $|\varphi\rangle$ are orthogonal; that is, their inner product $\langle\psi|\varphi\rangle=0$.
\end{itemize}
Let $S(\hs)$ denote the set of all subspaces of $\hs$. Then it is well-known that $(S(\hs),\wedge,\vee,\perp)$ is a complete modular lattice \cite{BvN36}. In this paper, given a quantum circuit $C$ with Hilbert space $\hs$, we specify the correctness of $C$ using the Hoare triple $\{X\}\ C\ \{Y\}$, where $X,Y \in S(\hs)$. This denotes that for any input state $\vert{}\psi\rangle \in X$, all corresponding output states will belong to $Y$.

 \subsection{Technical Lemmas}
 
 Birkhoff–von Neumann quantum logic, described in the previous subsection, is primarily concerned with logical relations and operations on subspaces of a fixed Hilbert space. However, when reasoning about quantum circuits, we need to understand how these logical relations and operations behave under quantum gates (i.e., unitary transformations). We first present a lemma that characterises this behaviour. Furthermore, compositional reasoning about quantum circuits requires us to derive the correctness properties of larger circuits, such as sequential and parallel compositions, from those of their components. To this end, we need to transfer logical relations and operations between the Hilbert spaces associated with individual components and those associated with larger circuits. We therefore establish two lifting lemmas that facilitate such transfers.

Let $U$ be an unitary operator on a Hilbert space $\hs$ and $X$ a subspace of $\hs$. Then we define the image of $X$ under $U$ as \begin{equation}\label{eq-image}U(X)=\{U|\psi\rangle: |\psi\rangle\in X\}.\end{equation} It is easy to show that $U(X)$ is also a subspace of $\hs$ with the same dimension as $X$. Since $U^{-1}$ is also an unitary operator, $U^{-1}(X)$ is well-defined. 

The first lemma asserts that unitary operators preserve quantum logical operations.

\begin{lem}\label{lem-qo} Let $U$ be an unitary operator on $\hs$. Then for any subspaces $X$ and $X_i$ of $\hs$, we have: \begin{align}\label{u-cap}&U\left(\bigcap_i X_i\right)=\bigcap_i U(X_i);\\ \label{u-vee}&U\left(\bigvee_i X_i\right)=\bigvee_i U(X_i);\\ \label{u-neg}&U\left(X^\perp\right)=(U(X))^\perp.
\end{align}
\end{lem} 
 
 The next lemma shows that tensoring with an auxiliary Hilbert space preserves the quantum logical operations on subspaces: intersection, join, and orthocomplement. This allows quantum logical reasoning to extend naturally to larger quantum circuits.
 
\begin{lem}\label{lem-cap}Let $X$ and $Y$ be subspaces of Hilbert space $\hs$. Then for any Hilbert space $\mathcal{K}$, we have: \begin{align}&(X\cap Y)\otimes\mathcal{K}=(X\otimes\mathcal{K})\cap(Y\otimes\mathcal{K}),\\ &(X\vee Y)\otimes\mathcal{K}=(X\otimes\mathcal{K})\vee (Y\otimes\mathcal{K}),\\ &(X^\perp)\otimes\mathcal{K}=(X\otimes\mathcal{K})^\perp\end{align} 
\end{lem}

The correctness of a quantum circuit can be characterised by the inclusion $U(X)\subseteq Y$, where $U$ is the unitary operator denoted by the circuit and $X$ and $Y$ are its precondition and postcondition subspaces, respectively. The following lemma shows that this inclusion is preserved when the system is extended with an external environment $\mathcal{K}$. This enables us to establish the correctness of larger quantum circuits by reasoning about their individual components.

\begin{lem}[Lifting]\label{lem-lift} Let $\hs, U$ and $X$ as in equation (\ref{eq-image}). Then for any Hilbert space $\mathcal{K}\neq\{0\}$ and for any subspace $Y$ of $\hs$, we have: \begin{equation}\label{eq-lift}U(X)\subseteq Y\Leftrightarrow \left(U\otimes I_\mathcal{K}\right)(X\otimes\mathcal{K})\subseteq Y\otimes \mathcal{K}\end{equation} where $I_\mathcal{K}$ is the identity operator on $\mathcal{K}$. 
\end{lem}

For readability, the proofs of the above lemmas are deferred to the appendix.

\section{Quantum Circuits and Their Parallel Execution}\label{sec-qc}

In this section, we first recall the formal definitions of the syntax and semantics of quantum circuits. Building on these definitions, we introduce a semantic model for the parallel execution of quantum circuits. This model will serve as the basis for reasoning about concurrency in quantum computing at the level of quantum circuits. The soundness of all proof rules developed in this paper will be established with respect to this semantic model.

\subsection{Syntax of Quantum Circuits}

The syntax and semantics of quantum circuits presented in this subsection and the next are standard in the literature. Our presentation follows \cite{Ying26}. We assume a set $\mathit{qVar}$ of qubit variables and a set $\mathcal{U}$ of basic quantum gates. Quantum circuits are generated by the following syntax:
\begin{equation}\label{syn-c}C::=U[\overline{q}]\ |\ C_1;C_2\end{equation}
where $U\in\mathcal{U}$ is a basic quantum gate and $\overline{q}$ is a sequence of qubit variables in $\mathit{qVar}$ such that if $U$ is an $n$-qubit gates, then $\overline{q}$ consists of $n$ qubits. 
Intuitively, $U[\overline{q}]$ means that quantum gate performing on quantum register $\overline{q}$, and $C_1;C_2$ denotes the sequential composition of $C_1$ and $C_2$. We use $\mathit{qv}(C)$ to denote the set of qubit variables occurring in a quantum circuit $C$.   

The inverse of a quantum circuit \(C\), denoted \(C^{-1}\), is the circuit that reverses the computation performed by \(C\). It is constructed by reversing the order of all gates in \(C\) and replacing each gate with its inverse.
More precisely, it is inductively defined as follows:\begin{enumerate}\item if $C=U[\overline{q}]$, then $C^{-1}=U^{-1}[\overline{q}]$; \item if $C=C_1;C_2$, then $C^{-1}=C_2^{-1};C_1^{-1}.$
\end{enumerate}

\subsection{Semantics of Quantum Circuits}
Let us first introduce several notations. For any qubit $q\in\mathit{qVar}$, we write $\hs_q$ for its Hilbert space. For any $Q\subseteq\mathit{qVar}$, we define: $$\hs_Q=\bigotimes_{q\in Q}\hs_q.$$ The identity operator on $\hs_Q$ is denoted by $I_Q$. For any quantum state $|\psi\rangle\in\hs_Q$, we define $\mathit{qv}(|\psi\rangle)=Q$ for the quantum variables of $|\psi\rangle$. For any $Q\subseteq Q^\prime\subseteq\mathit{qVar}$ and operator $U$ on $\hs_Q$, the cylindrical extension of $U$ on $Q^\prime$ is defined as the following operator on $\hs_{Q^\prime}$: $$U\uparrow Q^\prime=Q\otimes I_{Q^\prime\setminus Q}.$$ 

The operational semantics of a quantum circuit $C$ is modelled by the sequential execution of quantum gates in $C$. To describe the operational semantics precisely, we define a configuration to be a pair $(C,|\psi\rangle)$, where $C$ is a quantum circuit and $|\psi\rangle\in\hs_Q$ for some $\mathit{qv}(C)\subseteq Q\subseteq\mathit{qVar}$. Intuitively, $C$ is the circuit to be executed or the termination symbol $\downarrow$, and $|\psi\rangle$ is the current state of qubits. Then:
\begin{defn}\label{def-op-sem} The operational semantics of quantum circuits is defined by the following two transition rules:\begin{align*}(\mbox{UT})\ \ \ \frac{U|\psi\rangle=|\psi^\prime\rangle}{(U[\overline{q}],|\psi\rangle)\rightarrow(\downarrow,|\psi^\prime\rangle)}\qquad\qquad (\mbox{SC})\ \ \ \frac{(C_1,|\psi\rangle)\rightarrow (C_1^\prime,|\psi^\prime\rangle)}{(C_1;C_2,|\psi\rangle)\rightarrow (C_1^\prime;C_2,|\psi^\prime\rangle)}
\end{align*} In the rule (SC), we use the convention that $\downarrow;C_2=C_2$. 
\end{defn}

Here, (UT) stands for unitary transformation and (SC) for sequential composition. It should be noted that operator $U$ in the premise of the rule (UT) is actually an abbreviation of its cylindrical extension $U\uparrow\mathit{qv}(|\psi\rangle)$. Since later in this paper, we will consider parallel execution of quantum circuits, it should be particularly pointed out that a basic quantum gate $U$ in the transition rule (UT) is always treated as an atomic action and thus cannot be interrupted by other actions.    

\begin{defn}\label{def-denot} Let $\mathit{qv}(C)\subseteq Q\subseteq\mathit{qVar}$. Then the denotational semantics of a quantum circuit $C$ on variables $Q$ is defined as an unitary operator $\llbracket C\rrbracket_Q$ on $\hs_Q$ by induction on the structure of $C$:\begin{enumerate}\item if $C=U[\overline{q}]$, then $\llbracket C\rrbracket_Q=U\uparrow Q$;
\item if $C=C_1;C_2$, then $\llbracket C\rrbracket_Q=\llbracket C_2\rrbracket_Q \llbracket C_1\rrbracket_Q.$
\end{enumerate} \end{defn}

We often simply write $\llbracket C\rrbracket$ for $\llbracket C\rrbracket_Q$ when $Q$ can be recognised from the context. 

It is easy to see that the operational and denotational semantics are equivalent in the following sense: for any quantum circuit $C$ and for any $|\psi\rangle, |\psi^\prime\rangle\in\hs_{\mathit{qv}(C)}$, \begin{equation}\label{sem-equiv}\llbracket C\rrbracket|\psi\rangle=|\psi^\prime\rangle\ \mbox{iff}\ (C,|\psi\rangle)\rightarrow^\ast (\downarrow,|\psi^\prime\rangle)\end{equation} 
where $\rightarrow^\ast$ stands for the reflective and transitive closure of the transition relation $\rightarrow$ in Definition \ref{def-op-sem}. Also, we can show that $\llbracket C^{-1}\rrbracket=\llbracket C\rrbracket^{-1}$ for any quantum circuit $C$. 

For compositional reasoning, we often need to examine the behaviour of a quantum circuit in the presence of external systems. The following lemma shows how the operational and denotational semantics of a quantum circuit extend to this larger setting.

\begin{lem}[Lifting of Semantics]\label{lem-lift1} Let $\mathit{qv}(C)\subseteq Q_1\subseteq Q_2\subseteq\mathit{qVar}$. Then:\begin{enumerate}\item for any $|\psi_i\rangle\in\hs_{Q_1}$ and $|\varphi_i\rangle\in\hs_{Q_2\setminus Q_1}$, 
$$\frac{(C,|\psi_i\rangle)\rightarrow (C^\prime, |\psi_i^\prime)\ \mbox{for all}\ i}{\left(C,\sum_i\alpha_i|\psi_i\rangle|\varphi_i\rangle\right)\rightarrow \left(C^\prime,\sum_i\alpha_i|\psi_i^\prime\rangle|\varphi_i\rangle\right)}$$
\item $\llbracket C\rrbracket_{Q_2}=\left(\llbracket C\rrbracket_{Q_1}\right)\uparrow Q_2.$
\end{enumerate}\end{lem}

The above lemma can be easily proved by induction on the structure of quantum circuit $C$. 

\subsection{Atomic Regions}\label{subsec-atom}

With the preliminaries from the previous subsections in place, we now turn to the parallel execution of quantum circuits. As in classical computing, concurrency involving shared qubits can significantly complicate the behaviour of a parallel quantum circuit. A major reason is that its component circuits can interfere with one another by applying quantum gates to shared qubits. In classical parallel programming, atomic regions were introduced to reduce the possible points of interference: each atomic region executes without interruption by other components. In this subsection, we generalise this notion to quantum circuits.

\begin{defn}\begin{enumerate}\item If $C$ is a quantum circuit defined by the syntax (\ref{syn-c}), then $\langle C\rangle$ is called an atomic region.
\item A quantum circuit with atomic regions is a sequence $C=C_1;...;C_n$, where $n\geq 1$, and each $C_i$ $(1\leq i\leq n)$ is either a basic quantum gate $U[\overline{q}]$ or an atomic region.
\end{enumerate}\end{defn}

It is clear from the above definition that an atomic region cannot contain other atomic regions; that is, atomic regions are not allowed to be nested. 

The operational semantics of atomic regions is defined by the following transition rule:
$$(\mbox{AR})\ \ \ \frac{(C,|\psi\rangle)\rightarrow^\ast (\downarrow,|\psi^\prime)}{(\langle C\rangle,|\psi\rangle)\rightarrow (\downarrow,|\psi^\prime)}.$$ The transition rule (AR) reduces the multi-step computation of the body $C$ of an atomic region $\langle C\rangle$ into a single step and thus reflects the intuitive meaning of atomic region. Accordingly, the denotational semantics of atomic regions and quantum circuits with atomic regions are defined as follows:\begin{enumerate}\item $\llbracket\langle C\rangle \rrbracket=\llbracket C\rrbracket$;
\item If $C=C_1;...;C_n$ is a quantum circuit with atomic regions, then $\llbracket C\rrbracket=\llbracket C_n\rrbracket\cdots\llbracket C_1\rrbracket.$
\end{enumerate} It is easy to see that the equivalence (\ref{sem-equiv}) between operational and denotational semantics can be generalised to quantum circuits with atomic regions.  

\subsection{Parallel Execution of Quantum Circuits}

To specify parallel execution of quantum circuits, we introduce the following syntax for parallel compositions of quantum circuits:
\begin{equation}\label{syn-p}C::= C_1\|\cdots\|C_n\end{equation}
where $n\geq 2$, and $C_1,...,C_n$ are quantum circuits with atomic regions. Intuitively, $C_1\|\cdots\|C_n$ means that $n$ quantum circuits $C_1,...,C_n$ run in parallel. 

Let $C=C_1\|\cdots\|C_n$ be a parallel quantum circuit. Then the operational semantics of $C$ is defined by the interleaving execution model:\begin{itemize}\item \textit{ if one of its components, say $C_i$, performs an atomic action, then $C$ performs the same atomic action. Thus, the executions of two different components, say $C_i$ and $C_j$ $(i\neq j)$, may overlap, and their overlapping is modelled by executing them sequentially, in either order.}\end{itemize} Formally, we have:

\begin{defn}\label{def-pe}The operational semantics of parallel quantum circuit $C=C_1\|\cdots\|C_n$ is defined by the following transition rule: for any $1\leq i\leq n$ and for any $|\psi\rangle$ with $\mathit{qv}(|\psi\rangle)\supseteq\mathit{qv}(C)=\bigcup_{i=1}^n\mathit{qv}(C_i)$, 
\begin{equation*}(\mbox{PC})\ \ \ \frac{(C_i,|\psi\rangle)\rightarrow (C^\prime_i,|\psi^\prime\rangle)}{(C,|\psi\rangle)\rightarrow (C_1\|\cdots\|C_{i-1}\|C_i^\prime\|C_{i+1}\|\cdots\|C_n,|\psi^\prime\rangle)}
\end{equation*}\end{defn} Here, (PC) stands for parallel composition. We define the termination of $C$ as $\downarrow\ =\ \downarrow\|\cdots\|\downarrow$; that is, $C$ terminates iff all its components $C_i$ $(i=1,...,n)$ terminate. 

Note that the denotational semantics of a single quantum circuit $C$ is a unitary operator $\llbracket C\rrbracket$ on $\hs_Q$ for some $Q\supseteq \mathit{qv}(C)$, which maps each quantum state $|\psi\rangle$ to a quantum state $\llbracket C\rrbracket|\psi\rangle$. However, due to the nondeterminism introduced in the interleaving execution model, the denotational semantics of a parallel quantum circuit $C$ should be defined as a set-valued function; that is, for a single input, $C$ may have multiple possible outputs. More precisely, we have: 

\begin{defn}\label{def-p-denot}Let $\mathit{qv}(C)\subseteq Q\subseteq\mathit{qVar}$. The denotational semantics of a parallel quantum circuit $C$ on variables $Q$ is defined as the mapping from a quantum state in $\hs_Q$ to a set of quantum states in $\hs_Q$: for any $|\psi\rangle$ in $\hs_Q$, 
$$\llbracket C\rrbracket|\psi\rangle=\left\{|\psi^\prime\rangle:(C,|\psi\rangle)\rightarrow^\ast(\downarrow,|\psi^\prime\rangle)\right\}$$ where $\rightarrow^\ast$ stands for the reflexive and transitive closure of the transition relation $\rightarrow$ defined in Definition \ref{def-pe}. 
\end{defn}

To illustrate the above definitions, let us see a simple example. 

\begin{exam}Let us consider the parallel quantum circuit $P^\prime$ in Example \ref{exam-1}. According to the transition rule (PC) in Definition \ref{def-pe}, when executing $P^\prime$ on the input state $|1\rangle_{q_1}|0\rangle_{q_2}|0\rangle_{q_3}$, in addition to the final states $|+\rangle|1\rangle|0\rangle$, $|0\rangle|-\rangle|0\rangle$ and $\frac{1}{\sqrt{2}}(|10\rangle-|01\rangle)|0\rangle$ given in Example \ref{exam-1}, $P^\prime$ allows the following additional execution path: 
\begin{equation*}|1,0\rangle|0\rangle\stackrel{\mbox{CNOT}[q_1,q_2]}{\rightarrow}|1,1\rangle|0\rangle\stackrel{\mbox{CNOT}[q_2,q_1]}{\rightarrow}|0,1\rangle|0\rangle\stackrel{H[q_1]}{\rightarrow} |+,1\rangle|0\rangle\stackrel{\mbox{CNOT}[q_1,q_2]}{\rightarrow}\frac{1}{\sqrt{2}}(|01\rangle+|10\rangle)|0\rangle,\end{equation*} which yields another entangle final state. Therefore, we have:
\begin{align*}\llbracket P^\prime\rrbracket(|100\rangle)=\left\{|+10\rangle, |0-0\rangle, \frac{1}{\sqrt{2}}(|10\rangle\pm|01\rangle)|0\rangle\right\}. 
\end{align*} However, if we require $\mbox{SWAP}^\prime$ in $P^\prime$ to be executed atomically, that is, if we define $$P^{\prime\prime}\equiv\langle\mbox{SWAP}^\prime[q_1,q_2]\rangle\|H[q_1]$$ then it is easy to see that 
$$\llbracket P^{\prime\prime}\rrbracket(|100\rangle)=\llbracket P\rrbracket(|100\rangle)=\{|+10\rangle,|0-0\rangle\}.$$
\end{exam}

\section{Correctness Formulas}\label{sec-correct}

The operational and denotational semantics of parallel quantum circuits defined in the previous section provide a formal foundation for reasoning about shared-qubit concurrency in quantum computing. To facilitate such reasoning, our next goal is to develop effective proof rules for parallel quantum circuits. As a first step towards this goal, in this section, we introduce the notion of a correctness formula for formally specifying the correctness of quantum circuits.

For any set $Q\subseteq\mathit{qVar}$ of quantum variables, we adopt the idea of Birkhoff-von Neumann quantum logic that a proposition about the (composed) quantum system of quantum variables $Q$ is a subspace $X$ of Hilbert space $\hs_Q$. In this case, we define $\mathit{qv}(X)=Q$ for the quantum variables of $X$ (see Section \ref{sec-prel}). If $Q\subseteq Q^\prime\subseteq\mathit{qVar}$, and $X$ is a subspace of $\hs_Q$, then its cylindrical extension on $Q^\prime$ is defined to be the subspace of $\hs_{Q^\prime}$:
\begin{equation}\label{eq-cylindrical}X\uparrow Q^\prime=X\otimes\hs_{Q^\prime\setminus Q}.\end{equation}

There is a subtle difference between classical and quantum propositions. For example, $x=y$ is a proposition about classical variables $x$ and $y$. It can be freely seen as a proposition about variables $x,y$ and $z$ (with no restriction on $z$). However, if $Q\subsetneq Q^\prime\subseteq\mathit{qVar}$, then a proposition $X$ about $Q$ cannot be directly viewed as a proposition about $Q^\prime$. Instead, one needs to use its cylindrical extension $X\uparrow Q^\prime$ for this purpose.

Using subspaces of a Hilbert space as preconditions and postconditions, we can introduce:  

\begin{defn}\label{def-H}Let $C$ be a (parallel) quantum circuit with $\mathit{qv}(C)\subseteq Q\subseteq\mathit{qVar}$, and let $X,Y$ be subspaces of $\hs_Q$. Then \begin{equation}\label{eq-H}\{X\}\ C\ \{Y\}\end{equation} is called a correctness formula or Hoare triple for $C$.    
\end{defn}

Quantum Hoare triples were first introduced in \cite{DP06} for general precondition $X$ and postcondition $Y$ as observables modelled by Hermitian operators between the zero and identity operators. The special class of Hoare triples with subspaces as preconditions and postconditions as in Birkhoff-von Neumann quantum logic was studied in \cite{Ying10}.

Intuitively, $\{X\}\ C\ \{Y\}$ means that the output of quantum circuit $C$ lies in subspace $Y$ provided its input is from subspace $X$. Formally, we have:
\begin{defn} Hoare triple $\{X\}\ C\ \{Y\}$ is said to be valid, written $$\models\{X\}\ C\ \{Y\}$$ if for any $|\psi\rangle\in X$, it holds that $\llbracket C\rrbracket|\psi\rangle\subseteq Y;$ equivalently $\llbracket C\rrbracket(X)\subseteq Y,$ where $\llbracket C\rrbracket(X)=\bigcup_{|\psi\rangle\in X}\llbracket C\rrbracket|\psi\rangle.$
\end{defn}

It should be pointed out that the above definition applies to both sequential and parallel executions of quantum circuits. In the sequential case, it suffices to require that, for every $|\psi\rangle\in X$, we have $\llbracket C\rrbracket|\psi\rangle\in Y$.

To illustrate the above definition, let us see an example of parallel quantum circuits. 

 \begin{exam}\begin{enumerate}\item We first consider a disjoint parallel quantum circuit $C=H[q_1]\| H[q_2]$. Let $E=\mbox{span}\left\{|\beta_{00}\right\}$, where $|\beta_{00}\rangle=\frac{1}{\sqrt{2}}(|00\rangle+|11\rangle)$ is the EPR pair. Then it holds that $\models\{E\}\ C\ \{E\}$. 
 
 \item Let $A=\mbox{span}\left\{|+\rangle_{q_1}|0\rangle_{q_2},|0\rangle_{q_1}|1\rangle_{q_2}\right\}.$ Then we have:
 $$\{E\}\ \mbox{CNOT}[q_1,q_2]\|\mbox{CNOT}[q_2,q_1]\ \{A\}.$$
 
\item (Noncommuting exchange gates on an entangled subspace) 
Consider four qubits in the order $q_1q_2q_3q_4$, and define
\begin{align}
 |e_0\rangle
 &=\frac{1}{2}\bigl(
 |0101\rangle-|0110\rangle-|1001\rangle+|1010\rangle
 \bigr), \label{qcc-ex-e0}\\
 |e_1\rangle
 &=\frac{1}{\sqrt{3}}\left(
 |0011\rangle+|1100\rangle
 -\frac{|0101\rangle+|0110\rangle+|1001\rangle+|1010\rangle}{2}
 \right). \label{qcc-ex-e1}
\end{align}
These vectors are orthonormal. Let
$X=\operatorname{span}\{|e_0\rangle,|e_1\rangle\}$, $S$ stand for the SWAP operator, 
and 
\[
 R(\theta)=e^{-i\theta S}
              =\cos\theta\,I-i\sin\theta\,S.
\]
We set the parallel quantum circuit: \[
 P_{\theta,\phi}
 =R(\theta)[q_1,q_2]\ \|\ R(\phi)[q_2,q_3].
\]
Here, $R(\theta)$ is treated as a basic
gate, hence as an atomic action at the chosen level of abstraction, and the two components share $q_2$. It can be checked that for all real $\theta,\phi$, we have: \begin{equation}
 \models\{X\}\ P_{\theta,\phi}\ \{X\}.
 \label{qcc-ex-code-invariant}
\end{equation}\end{enumerate}
\end{exam}

The following lemma shows that the correctness of a quantum circuit is preserved when it is considered in a larger environment. As we will see shortly, this lemma, together with the proof rule (R-SC) to be presented in Figure \ref{fig-proof1}, enables us to conveniently reason about the correctness of sequentially composed quantum circuits acting on different sets of qubits.

\begin{lem}[Lifting of Correctness]\label{lem-clift} Let $\mathit{qv}(C)\subseteq Q_1\subseteq Q_2\subseteq\mathit{qVar}$, and let $X,Y$ be subspaces of $\hs_{Q_1}$. Then
\begin{equation}\models\{X\}\ C\ \{Y\}\Leftrightarrow \{X\uparrow Q_2\}\ C\ \{Y\uparrow Q_2\}.\end{equation}
\end{lem}
\begin{proof} Immediate from Lemmas \ref{lem-lift} and \ref{lem-lift1}(2). 
\end{proof}

\section{Proof Rules for Sequential Execution of Quantum Circuits}\label{sec-proof-qc1}

Using the notion of a correctness formula introduced above, we present a series of proof rules for reasoning about the sequential execution of quantum circuits in this section. Conceptually, these rules serve as a stepping stone towards our ultimate goal of reasoning about parallel quantum circuits. In practical applications, they will be used in combination with the proof rules for parallel execution to be developed in Section \ref{sec-proof-qc2}.

A proof rule is of the form $$\frac{T_1,\ldots,T_n}{T},$$ 
where $T_1,\ldots,T_n$ and $T$ are all correctness formulas. It means that the premises $T_1,\ldots,T_n$ entail the conclusion $T$. In particular, when $n=0$, the rule has no premises, and $T$ is called an axiom. The basic proof rules for the sequential execution of quantum circuits are presented in Figure \ref{fig-proof1}. These rules are essentially special cases of quantum Hoare logic \cite{Ying11}, tailored to quantum circuits and restricted to subspaces as preconditions and postconditions, as in \cite{Zhou19}.
\begin{figure}[h]\centering
\begin{align*}&(\mbox{R-UT})\ \ \ \left\{U^{-1}(X)\right\}\ U[\overline{q}]\ \{X\}\\ 
&\mbox{(R-SC)}\ \ \ \frac{\{X\}\ C_1\ \{W\}\qquad \{Y\}\ C_2\ \{Z\}\qquad W\uparrow\left(\mathit{qv}(W)\cup\mathit{qv}(Y)\right)\subseteq Y\uparrow\left(\mathit{qv}(W)\cup\mathit{qv}(Y)\right)}{\left\{X\uparrow\left(\mathit{qv}(W)\cup\mathit{qv}(Y)\right)\right\}\ C_1;C_2\ \left\{Z\uparrow\left(\mathit{qv}(W)\cup\mathit{qv}(Y)\right)\right\}}\\ &(\mbox{R-}\neg)\ \ \ \frac{\{X\}\ C\ \{Y\}}{\left\{Y^\perp\right\}\ C^{-1}\ \left\{X^\perp\right\}} 
\\ &(\mbox{R-}\subseteq)\ \ \ \frac{X\subseteq X^\prime\qquad \{X^\prime\}\ C\ \{Y^\prime\}\qquad Y^\prime\subseteq Y}{\{X\}\ C\ \{Y\}} 
\\ &(\mbox{R-}\cap)\ \ \ \frac{\{X_i\}\ C\ \{Y_i\}\ \mbox{for}\ i=1,...,n}{\left\{\bigcap_{i=1}^n X_i\right\}\ C\ \left\{\bigcap_{i=1}^n Y_i\right\}}\qquad\qquad (\mbox{R-}\vee)\ \ \ \frac{\{X_i\}\ C\ \{Y_i\}\ \mbox{for}\ i=1,...,n}{\left\{\bigvee_{i=1}^n X_i\right\}\ C\ \left\{\bigvee_{i=1}^n Y_i\right\}}\end{align*}
\caption{Proof Rules for Sequential Execution of Quantum Circuits}\label{fig-proof1}
\end{figure}

The rule (R-UT) asserts that the output state of a quantum gate $U[\overline{q}]$ lies in a subspace $X$, provided that its input state lies in $U^{-1}(X)$. The rule (R-SC) allows us to infer the correctness of a sequential composition $C_1;C_2$ from the correctness of its component circuits $C_1$ and $C_2$. It is worth noting that this rule differs from its counterparts in \cite{Ying11, Zhou19}. In Figure \ref{fig-proof1}, the postcondition $W$ of $C_1$ and the precondition $Y$ of $C_2$ are allowed to involve different sets of qubits; that is, $\mathit{qv}(W)\neq\mathit{qv}(Y)$. Consequently, a condition relating $W$ and $Y$ is added as the third premise of (R-SC). Fortunately, this condition is usually easy to check in practical applications, where the sets of qubits involved in $C_1$ and $C_2$ typically differ only slightly. The rule ($\mbox{R-}\neg$) allows us to infer the correctness of the inverse of a quantum circuit from the correctness of the circuit itself. The rule ($\mbox{R-}\subseteq$) enables us to strengthen the precondition and weaken the postcondition. Finally, the rules ($\mbox{R-}\cap$) and ($\mbox{R-}\vee$) allow us to apply the quantum logical operations of intersection and join to preconditions and postconditions.

The following example illustrates how these rules can be used to prove correctness of quantum circuits. 

\begin{exam}[Reusing a clean ancilla for a non-Clifford entangling circuit]
\label{qcc-ex-parity-phase}
Let $q_1,q_2,q_3$ be data qubits and $q$ a work qubit, and \[
 C_\theta=C_{12}(\theta);C_{23}(\theta);C_{13}(\theta).
\]
where 
\begin{align*} C_{ij}(\theta)
 \equiv D_{ij};R_z(2\theta)[a];D_{ij}^{-1},\qquad 
 D_{ij}
 \equiv\operatorname{CNOT}[q_i,a];
         \operatorname{CNOT}[q_j,a], \notag
\end{align*}
and $R_z(2\theta)=e^{-i\theta Z}$. The work qubit $q$ is initially in
$|0\rangle$. 
Let $$E_0^{ij}=\operatorname{span}\{|00\rangle,|11\rangle\},\qquad\qquad E_1^{ij}=\operatorname{span}\{|01\rangle,|10\rangle\}$$ 
be the even- and odd-parity subspaces of $q_iq_j$. For $b\in\{0,1\}$,
put $$L_b^{ij}=E_b^{ij}\otimes\operatorname{span}\{|0\rangle_q\},\qquad\qquad M_b^{ij}=E_b^{ij}\otimes\operatorname{span}\{|b\rangle_q\}.$$ 
The gate rule (R-UT), followed by the sequential composition rule (R-SC), gives
\begin{equation}\vdash 
 \{L_b^{ij}\}\ D_{ij}\ \{M_b^{ij}\},
 \qquad
 \vdash \{M_b^{ij}\}\ R_z(2\theta)[a]\ \{M_b^{ij}\},
 \qquad
 \vdash \{M_b^{ij}\}\ D_{ij}^{-1}\ \{L_b^{ij}\}.
 \label{qcc-ex-parity-triples}
\end{equation}
Thus, (R-SC) gives $$\vdash \{L_b^{ij}\}C_{ij}(\theta)\{L_b^{ij}\}.$$
Applying (R-$\vee$) to $b=0,1$, we obtain
\begin{equation}
\vdash \{A_{ij}\}\ C_{ij}(\theta)\ \{A_{ij}\},
 \qquad
 A_{ij}=\mathcal{H}_{\{q_i,q_j\}}
             \otimes\operatorname{span}\{|0\rangle_q\}.
 \label{qcc-ex-clean-module}
\end{equation} Note that each $A_{ij}$ is declared on $\{q_i,q_j,a\}$. With
$Q=\{q_1,q_2,q_3,a\}$ and $$A=\mathcal{H}_{\{q_1,q_2,q_3\}}
      \otimes\operatorname{span}\{|0\rangle_q\},$$ the matching conditions in (R-SC) are exactly
\begin{equation}
 A_{12}\uparrow Q=A_{23}\uparrow Q=A_{13}\uparrow Q
 =A. 
 \label{qcc-ex-changing-registers}
\end{equation}
Consequently, repeated applications of (R-SC) yield: $$\vdash \{A\}\ C_\theta\ \{A\}.$$ 
This indicates that the same work qubit $q$ is safely reused across three modules
acting on different pairs of data qubits.
\end{exam}

To conclude this section, we present the following: 

\begin{thm}[Soundness]\label{thm-sound} All proof rules in Figure \ref{fig-proof1} are sound. 
\end{thm}
\begin{proof} See Appendix \ref{app-a}.
\end{proof}

\section{Proofs for Quantum Circuits with Atomic Regions}\label{sec-atom}

As in the Owicki-Gries method for reasoning about classical shared-memory concurrency \cite{OG76}, we need introduce the notion of proof for quantum circuits with atomic regions upon which interference freedom can be defined. First of all, we introduce a proof rule (R-AR) for atomic regions in Figure \ref{fig -atom}. 
\begin{figure}[h]\centering
\begin{equation*}\mbox{(R-AR)}\ \ \ \frac{\{X\}\ C\ \{Y\}}{\{X\}\ \langle C\rangle\ \{Y\}}\end{equation*}
\caption{Proof Rule for Atomic Regions.}\label{fig -atom}
\end{figure}
The soundness of the proof rule (R-AR) immediately follows from the transition rule (AR) in Subsection \ref{subsec-atom}. 

The only difference between the correctness proofs in the classical and quantum cases is that in the quantum case we need pay special attention on matching the preconditions and postcondition in the composition of a sequence of quantum circuits (see the condition (3) in the next definition).   

\begin{defn}\label{def-outline}Let $C=C_1;\cdots;C_n$, where each $C_i$ $(i=1,...,n)$ is either a basic quantum gate or an atomic region. Then a proof of correctness $\{X\}\ C\ \{Y\}$ is a sequence of the form:
\begin{equation}\label{eq-outline}\mathcal{P}=\{X_1\}\ C_1\ \{Y_1\}; \{X_2\}\ C_2\ \{Y_2\}; \cdots; \{X_n\}\ C_n\ \{Y_n\}\end{equation} such that \begin{enumerate}\item $X\subseteq X_1\uparrow\mathit{qv}(X)$ and $Y_n\uparrow\mathit{qv}(Y)\subseteq Y$; 
\item $\models\{X_i\}\ C_i\ \{Y_i\}$ for $i=1,...,n$; and 
\item $Y_i\uparrow\left(\mathit{qv}(Y_i)\cup\mathit{qv}(X_{i+1})\right)\subseteq X_{i+1}\uparrow\left(\mathit{qv}(Y_i)\cup\mathit{qv}(X_{i+1})\right)$ for $i=1,...,n-1$. 
\end{enumerate}
\end{defn}

We write: $$C[i)=C_i;C_{j+1};\cdots;C_n$$ for the tail of circuit $C$, and $\mathit{pre}(C[i))=X_i$ for the precondition of $C[i)$ in the proof (\ref{eq-outline}) for $i=1,...,n$.

It is particularly worth pointing out that in the proof (\ref{eq-outline}), for each $i=1,...,n-1$, the postcondition $Y_i$ of sub-circuit $C_i$ can be different from the precondition $X_{i+1}$ of $C_{i+1}$. Indeed, it is even allowed that $\mathit{qv}(Y_i)\neq\mathit{qv}(X_{i+1})$. This flexibility comes from our new proof rule (R-SC) for sequential composition (see the remark after equation (\ref{eq-cylindrical})). It is obvious that the condition (3) in the above definition is a reflection of the third condition in the premise of (R-SC). An important benefit of this flexibility is that in proving the correctness $\{X\}\ C\ \{Y\}$, we can make each step $\{X_i\}\ C_i\ \{Y_i\}$ local in the sense that only necessary variables, namely those occurring in $C_i$, are involved. In particular, whenever $C_i$ is a basic quantum gates which usually only have one or two qubits, checking the correctness $\{X_i\}\ C_i\ \{Y_i\}$ becomes very easy.

The soundness of proof rules (Theorem \ref{thm-sound}) can be generalised to proofs of quantum circuits with atomic regions: 

\begin{thm}\label{thm-sound-atom}If a correctness formula $\{X\}\ C\ \{Y\}$ has a proof, then it holds that $\models \{X\}\ C\ \{Y\}$. 
\end{thm}
\begin{proof} See Appendix \ref{app-b}.\end{proof}

\section{Proof Rules for Parallel Execution of Quantum Circuits}\label{sec-proof-qc2}

In this section, we establish a proof rule for compositional reasoning about parallel quantum circuits. As in the classical Owicki–Gries method, its premises combine local correctness with preservation of the assertions used by other components.

\subsection{Disjoint Parallel Quantum Circuits}\label{sec-disjoint}

Although our ultimate goal is to reason about shared-qubit concurrency, as a warm-up, let us first consider the special case of disjoint parallel quantum circuits: 
\[
C=C_1\|\cdots \|C_n,
\]
where $\mathit{qv}(C_i)\cap\mathit{qv}(C_j)=\emptyset\) for all \(1\leq i\neq j\leq n$. Suppose further that $\mathit{qv}(X_i)=\mathit{qv}(Y_i)=\mathit{qv}(C_i)\) for all \(i=1,\ldots,n$. Under these assumptions, the proof rule (R-PC-DQ), shown in Figure \(\ref{fig -PC-DQ}\), was established in \cite{Ying22}.
\begin{figure}[h]\centering
\begin{align*}\mbox{(R-PC-DQ)}\ \ \ \frac{\{X_i\}\ C_i\ \{Y_i\}\ (i=1,...,n)}{\left\{\bigotimes_{i=1}^nX_i\right\}\ C_1\|\cdots\|C_n\ \left\{\bigotimes_{i=1}^nY_i\right\}}
\end{align*}
\caption{Proof Rule for Parallel Execution of Quantum Circuits.}\label{fig -PC-DQ}
\end{figure}

Due to the disjointness condition, tensor products $\bigotimes$ can be used to represent logical conjunctions in the precondition and postcondition of $C_1\|\cdots \|C_n$ in the conclusion of the rule (R-PC-DQ). However, this approach is no longer applicable when the component circuits $C_1,\ldots,C_n$ share qubits.

\subsection{Conjunction of Quantum Propositions}

To specify the preconditions and postconditions of shared-qubit parallel quantum circuits in our proof rule, we need to generalise the quantum logical conjunction $\cap$, originally defined for subspaces of the same Hilbert space, to subspaces of different Hilbert spaces. This generalisation is necessary because different components of a parallel quantum circuit may involve different sets of qubit variables.

\begin{defn}\label{def-conj} For each $1\leq i\leq n$, let $X_i$ be a subspace of Hilbert space $\hs_{Q_i}$, where $Q_i\subseteq\mathit{qVar}$. We set $Q=\bigcup_{i=1}^nQ_i$. Then the conjunction of $X_i$ $(i=1,...,n)$ is defined to be the following subspace of $\hs_Q$:
\begin{equation}\label{def-ext-cap}\bigwedge_{i=1}^n X_i=\bigcap_{i=1}^n\left(X_i\uparrow Q\right).
\end{equation}\end{defn}

Intuitively, equational (\ref{def-ext-cap}) means that the conjunction holds precisely when all its conjuncts hold after
extension to a common register. If $Q\subseteq R$, Lemma \ref{lem-cap} gives
\[
 \left(\bigwedge_{i=1}^{n}X_i\right)\uparrow R
 =\bigcap_{i=1}^{n}(X_i\uparrow R).
\]
In particular, if $\mathit{qv}(X_i)=Q_i$ for every $i=1,...,n$ and $Q_i\cap Q_j=\emptyset$ for any $1\leq i\neq j\leq n$, then it is easy to check that \begin{equation}\label{eq-prod-cap}\bigwedge_{i=1}^nX_i=\bigotimes_{i=1}^nX_i.\end{equation}

\subsection{Interference Freedom}

Using the conjunction introduced in Definition \ref{def-conj}, we can now define the notion of interference freedom for quantum circuits.

Let us first define the notion of invariance of a subspace (viewed as an assertion) under the execution of a quantum circuit, subject to a given precondition.
\begin{defn}Let $C$ be a quantum circuit and $X$ a subspace of $\hs_{\mathit{qv}(C)}$, and let $Y$ be a subspace of $\hs_Q$, where $Q\subseteq\mathit{qVar}$. If the Hoare triple: $$\models\{X\wedge Y\}\ C\ \{Y\}$$ is valid, then we say that $Y$ is invariant under $C$ with precondition $X$. 
\end{defn}

Then interference freedom in a parallel quantum circuit requires that every assertion in the local proof of each component circuit be invariant under the execution of any atomic step of every other component circuit. This invariance must hold at every control location, including the terminal location of a completed component.

\begin{defn}\label{def-nonint}For each $i=1,...,n$, let $C_i=C_{i1};\cdots;C_{in_i}$ be a quantum circuit and $$\mathcal{P}_i=\{X_{i1}\}\ C_{i1}\ \{Y_{i1}\}; \{X_{i2}\}\ C_{i2}\ \{Y_{i2}\}; \cdots; \{X_{in_i}\}\ C_{in_i}\ \{Y_{in_i}\}$$ be a proof of correctness formula $\{X_i\}\ C_i\ \{Y_i\}$. If for any $1\leq i,j\leq n$ with $i\neq j$, and for any $1\leq k\leq n_i$ and $1\leq l\leq n_j$, $X_{ik}$ is invariant under $C_{jl}$ with $\mathit{pre}(C_{jl})=X_{jl}$, then proofs $\mathcal{P}_i\ (i=1,...,n)$ are said to be interference free. 
\end{defn}

To illustrate the above definitions, let us consider a simple example of two local proofs that interfere with each other.
\begin{exam} Consider the parallel quantum circuit:
$$\operatorname{CNOT}[q_1,q_2]\parallel H[q_1].$$
Let $X=\operatorname{span}\{|1\rangle\},$ $Y=\operatorname{span}\{|-\rangle\}$, $W=\operatorname{span}\{|10\rangle\},$ and $Z=\operatorname{span}\{|11\rangle\}.$
By the proof rule (R-UT) in Figure \ref{fig-proof1}, we can prove that the following two local Hoare triples
$$\vdash \{W\}\operatorname{CNOT}[q_1,q_2]\{Z\},
 \qquad
\vdash \{X\}H[q_1]\{Y\}$$
are valid, but their proofs are not interference free. The required
preservation of $W$ by $H[q_1]$ with precondition $X$ fails:
$W\cap(X\uparrow\{q_1,q_2\})=W$, while
\[
 H[q_1]|10\rangle
 =\frac{|00\rangle-|10\rangle}{\sqrt2}\notin W.
\]
Therefore the proof rule (R-PC-Q) to be introduced in the next subsection cannot be applied to these proofs.
\end{exam}

\subsection{The Parallel Rule}

We are now ready to present the proof rule (R-PC-Q) for reasoning about the parallel execution of quantum circuits, as shown in Figure \ref{fig -PC-Q}. It should be noted that, in the premises of rule (R-PC-Q), the preconditions $X_i$ of different component circuits $C_i$ may involve different sets of qubit variables. The conjunctions appearing in the precondition and postcondition of the parallel circuit $C_1\|\cdots \|C_n$ in the conclusion of the rule are defined according to Definition \ref{def-conj}.  

\begin{figure}[h]\centering
\begin{align*}\mbox{(R-PC-Q)}\ \ \ \frac{\mbox{Proofs of}\ \{X_i\}\ C_i\ \{Y_i\}\ (i=1,...,n)\ \mbox{are interference free}}{\left\{\bigwedge_{i=1}^nX_i\right\}\ C_1\|\cdots\|C_n\ \left\{\bigwedge_{i=1}^nY_i\right\}}
\end{align*}
\caption{Proof Rule for Parallel Execution of Quantum Circuits.}\label{fig -PC-Q}
\end{figure}

It is interesting to note that, in the case of disjoint parallelism considered in Subsection \ref{sec-disjoint}, interference freedom holds automatically. Consequently, by equation (\ref{eq-prod-cap}), the proof rule (R-PC-Q) reduces to the rule (R-PC-DQ).

An application of rule (R-PC-Q) to reasoning about error syndromes in quantum error correction will be presented as a case study in the next section. 

To conclude this section, we present the following: 
\begin{thm}\label{thm-sound-parallel}The proof rule (R-PC-Q) is sound.
\end{thm}
\begin{proof} See Appendix \ref{app-c}. 
\end{proof}

\section{Case Study}\label{sec-case}

In this section, we present an example to show how the proof rule (R-PC-Q) can be used to reason about coherent syndrome extraction during concurrent logical processing. 
 
\begin{exam} 
Let $D=\{q_1,q_2,q_3,q_4\}$ be a shared data register, and let $a,b$ be
private work qubits of the two components. Write $X_k,Z_k$ for Pauli
operators on $q_k$, and define
\begin{equation}
 S_Z=Z_1Z_2Z_3Z_4,\qquad S_X=X_1X_2X_3X_4,\qquad
 \overline X=X_1X_2,\qquad \overline Z=Z_2Z_3.
 \label{qcc-s7-paulis}
\end{equation}
The two stabilizers $S_Z,S_X$ commute, and both commute with
$\overline X,\overline Z$. However, $\overline X$ and $\overline Z$ anti-commute, that is, 
$\overline X\overline Z=-\overline Z\overline X$.
The code subspace
\[
 \mathcal C=\ker(S_Z-I)\cap\ker(S_X-I)
\]
has dimension four. Let $L_X(\alpha)=e^{-i\alpha X\otimes X}$ and
$L_Z(\beta)=e^{-i\beta Z\otimes Z}$ be two-qubit basic gates.
On $q_1q_2$ and $q_2q_3$, respectively, their cylindrical extensions are
$e^{-i\alpha\overline X}$ and $e^{-i\beta\overline Z}$.
These extensions preserve $\mathcal C$ and generally do not commute.
We use the same gate symbols for these extensions when the register is clear. For $ij\in\{12,34\}$, we define the two-gate atomic regions: 
\begin{equation*}
 A_{ij}=\langle\operatorname{CZ}[a,q_i];
                    \operatorname{CZ}[a,q_j]\rangle, \qquad
 B_{ij}=\langle\operatorname{CNOT}[b,q_i];
                    \operatorname{CNOT}[b,q_j]\rangle.
\end{equation*}
Consider the parallel execution $C_A\ \|\ C_B$ of the following two components: 
\begin{align}\label{eq-CA}
 C_A&=H[a];A_{12};A_{34};H[a];L_X(\alpha)[q_1,q_2],\\
 \label{eq-CB}C_B&=H[b];B_{12};B_{34};H[b];L_Z(\beta)[q_2,q_3].
\end{align}
The first component coherently records the eigenvalue of $S_Z$ in $a$,
and the second records that of $S_X$ in $b$. They also perform possibly
non-Clifford logical rotations while accessing the same data register.\end{exam}

\paragraph{\textbf{Notations.}} Let $\mathcal E_+(M)=\ker(M-I)$ for a Hermitian unitary $M$.
We annotate each component after $k$ of its five atomic steps by
$P_{A,k}=\mathcal E_+(M_{A,k})$ and
$P_{B,k}=\mathcal E_+(M_{B,k})$, where
\begin{equation}
\begin{array}{c|c|c}
 k&M_{A,k}&M_{B,k}\\ \hline
 0&Z_a&Z_b\\
 1&X_a&X_b\\
 2&X_aZ_1Z_2&X_bX_1X_2\\
 3&X_aS_Z&X_bS_X\\
 4&Z_aS_Z&Z_bS_X\\
 5&Z_aS_Z&Z_bS_X
\end{array}
 \label{qcc-s7-assertion-table}
\end{equation}
The $A$ assertions are declared on $D\cup\{a\}$ and the $B$ assertions
on $D\cup\{b\}$. Comparisons between components use their cylindrical
extensions to $R=D\cup\{a,b\}$.

\paragraph{\textbf{Local correctness.}} We first prove the correctness of component circuits $C_A$ and $C_B$. 
For a controlled involution $T$, conjugation of the control's Pauli $X$
gives $X_{\mathrm{control}}T$. Consequently,
\[
 A_{12}X_aA_{12}^\dagger=X_aZ_1Z_2,\qquad
 B_{12}X_bB_{12}^\dagger=X_bX_1X_2.
\]
The second paired region adds the remaining two Pauli factors. The
Hadamard gates exchange $X$ and $Z$ on their private work qubits.
Finally, $L_X(\alpha)$ commutes with $Z_aS_Z$, and
$L_Z(\beta)$ commutes with $Z_bS_X$.
Thus, for the $k$th atomic action $G_{A,k}$ or $G_{B,k}$, we have:
\[
 G_{A,k}M_{A,k-1}G_{A,k}^\dagger=M_{A,k},
 \qquad
 G_{B,k}M_{B,k-1}G_{B,k}^\dagger=M_{B,k}.
\]
The gate rule (R-UT) and atomic-region rule (R-AR) give the proofs: 
\begin{align}\label{eq-proof-CA}
 &\{P_{A,0}\}\ G_{A,1}\ \{P_{A,1}\};\cdots;
 \{P_{A,4}\}\ G_{A,5}\ \{P_{A,5}\},\\  
 \label{eq-proof-CB}
 &\{P_{B,0}\}\ G_{B,1}\ \{P_{B,1}\};\cdots;
 \{P_{B,4}\}\ G_{B,5}\ \{P_{B,5}\}. \end{align}

\paragraph{\textbf{Interference freedom.}} Now we check that proofs (\ref{eq-proof-CA}) and (\ref{eq-proof-CB}) do not interfere with each other. 
Every atomic action of $C_B$ commutes with every observable $M_{A,k}$,
and every atomic action of $C_A$ commutes with every $M_{B,k}$.
For the paired regions, the substantive identities are
\[
 [Z_1Z_2,X_1X_2]=0,\qquad [Z_3Z_4,X_3X_4]=0,
\]
each resulting from two Pauli anticommutations; regions on distinct pairs
also commute. The logical $X$ rotation commutes with the $X$-type data
operators in the $B$ assertions, and the logical $Z$ rotation commutes
with the $Z$-type data operators in the $A$ assertions.
Hence, for every $k$ and every action $G_{B,\ell}$,
\[
 G_{B,\ell}(P_{A,k}\uparrow R)=P_{A,k}\uparrow R,
\]
and symmetrically. These unconditional preservation identities imply all
the conditional obligations in Definition \ref{def-nonint}.
They include preservation of the terminal assertions at $k=5$.
There are $60$ directed checks when the six control locations of each
component are counted; every guard intersection has dimension $16$
in the $64$-dimensional global space. None is vacuous.

\paragraph{\textbf{Application of (R-PC-Q).}}
The conjunction of the initial assertions is
$\mathcal H_D\otimes\operatorname{span}\{|00\rangle_{ab}\}$.
The rule therefore gives
\begin{equation}\label{eq-syndrome}
\vdash \left\{\mathcal H_D\otimes\operatorname{span}\{|00\rangle_{ab}\}\right\}
 C_A\parallel C_B
 \left\{\mathcal Y\right\}\end{equation} where the postcondition: $$ \mathcal Y=\mathcal E_+(Z_aS_Z)\cap\mathcal E_+(Z_bS_X).$$
This holds for every input pure quantum state of the data and every
interleaving of the ten atomic steps. There are
$\binom{10}{5}=252$ such interleavings. Lifting correctness also permits
the data to be entangled with an arbitrary external reference register.

\paragraph{\textbf{Explanations.}} The postcondition in Hoare triple (\ref{eq-syndrome}) expresses correct syndrome correlations. More explicitly,
for $s,t\in\{+1,-1\}$, let
\[
 \Pi_{s,t}=\frac14(I+sS_Z)(I+tS_X),\qquad
 \epsilon(s)=\frac{1-s}{2}.
\]
Then
\begin{equation}
 \mathcal Y=
 \bigoplus_{s,t\in\{+1,-1\}}
 \operatorname{ran}(\Pi_{s,t})\otimes
 \operatorname{span}\{|\epsilon(s)\rangle_a|\epsilon(t)\rangle_b\}.
 \label{qcc-s7-syndrome-sectors}
\end{equation}
This allows coherent superpositions of syndrome sectors and entanglement
between the data and the work registers. For instance, on data input
$|0+0+\rangle$, direct evaluation gives
$\rho_{ab}=I_{ab}/4$ for every permitted schedule. The final global
pure quantum state has Schmidt rank four across $D\,|\,ab$.

The rule does not require the logical rotations to commute. Intersecting 
every assertion in \eqref{qcc-s7-assertion-table} with the lifted code
subspace $\mathcal C$ gives another interference-free pair of outlines,
and hence
\begin{equation}
 \{\mathcal C\otimes\operatorname{span}\{|00\rangle_{ab}\}\}\
 C_A\parallel C_B\
 \{\mathcal C\otimes\operatorname{span}\{|00\rangle_{ab}\}\}.
 \label{qcc-s7-code-safety}
\end{equation}
To exhibit schedule dependence, define
\[
 |0_L\rangle=\frac{|0000\rangle+|1111\rangle}{\sqrt2},
 \qquad
 |1_L\rangle=\frac{|1100\rangle+|0011\rangle}{\sqrt2}.
\]
On their span, $\overline X,\overline Z$ act as logical Pauli $X,Z$.
At $\alpha=\beta=\pi/4$, input $|0_L\rangle|00\rangle_{ab}$ has,
up to a common global phase, the two possible outputs
\begin{equation}
 \frac{|0_L\rangle+|1_L\rangle}{\sqrt2}|00\rangle_{ab},
 \qquad
 \frac{|0_L\rangle-i|1_L\rangle}{\sqrt2}|00\rangle_{ab}.
 \label{qcc-s7-two-outputs}
\end{equation}
Both satisfy the same correctness property proved by (R-PC-Q).

The paired atomic regions are essential to this proof. Although
$S_Z$ and $S_X$ commute, individual
$\operatorname{CZ}[a,q_i]$ and $\operatorname{CNOT}[b,q_i]$
do not commute. Exposing their internal gates permits schedules
that introduce an unwanted $\operatorname{CZ}[a,b]$ into the syndrome
extraction. On a code input, such a schedule can leave the work qubits in
$(|00\rangle+|01\rangle+|10\rangle-|11\rangle)/2$, violating
\eqref{eq-syndrome}.

\section{Related Work}\label{sec-related}

In this section, we discuss related work in order to accomodate the research described in this paper into the appropriate context.   

\paragraph{\textbf{Parallel quantum programming.}} Multiprogramming techniques were first introduced by Das et al. \cite{Das19} for quantum computers to enable noisy intermediate-scale quantum (NISQ) devices to execute multiple workloads concurrently, thereby improving their throughput and resource utilisation. An extension of the quantum programming language Q\# was proposed by  H\"{a}ner et al. \cite{Haner22} to support parallel quantum programming. Operational and denotational semantics for parallel quantum programs were developed in \cite{Ying18, Ying22}. A quantum programming language with explicit primitives for parallelism and synchronisation was introduced by Jian et al. \cite{Jain25}, together with a formal definition of its semantics. 
However, the proof system presented in \cite{Ying18, Ying22} was restricted to disjoint parallel quantum programs, whose components do not share quantum variables. In contrast, this paper develops an effective proof rule for reasoning about parallel quantum circuits with shared qubits.

\paragraph{\textbf{Quantum operating systems.}} Concurrency is a central concern in the design of operating systems \cite{textOS}. Several proposals for quantum operating systems have been reported in the literature \cite{QOS0, qinternet, QOS1, QOS2, Tao, QOS-Ye}. For example, Giortamis et al. \cite{QOS1} introduced a modular quantum operating system, QOS, that addresses the challenges of quantum resource management through a systematic exploration of key design trade-offs across the system stack. Ye, Wang, Qiao, and Palsberg \cite{QOS-Ye} proposed HALO, the first quantum operating system designed to support fine-grained resource sharing. Tao et al. \cite{Tao} developed HyperQ, a system that introduces virtual machines for quantum computers to provide fault isolation, improve resource utilisation, and reduce latency in quantum cloud computing. Delle Donne et al. \cite{qinternet} implemented QNodeOS, a quantum network operating system. However, these works have not systematically examined shared-memory concurrency in quantum computing. We expect that the work presented in this paper can be extended to provide a foundation for reasoning about concurrency in quantum operating systems. 

\section{Conclusion}\label{sec-con} In this paper, we have generalised the Owicki–Gries method for reasoning about shared-memory parallel programs to shared-qubit parallel quantum circuits. The key idea is to use subspaces of the Hilbert space of each component circuit as assertions and to define their conjunction as the intersection of their cylindrical extensions to the Hilbert space of the entire parallel circuit.

We believe that this approach could also be extended to generalise the rely–guarantee method \cite{Jones} for reasoning about shared-qubit parallel quantum circuits using subspaces as preconditions and postconditions. However, as noted in the Introduction, the approach does not directly extend to general parallel quantum programs that use arbitrary Hermitian operators (i.e., physical observables) as assertions, because a suitable notion of conjunction for such assertions remains to be developed. In our view, resolving this problem is a crucial step towards establishing a general theory of shared-memory quantum concurrency.

\smallskip\

\textbf{Acknowledgement:}  
This work was supported in part by the Australian Research Council under Grant No. DP250102952. ChatGPT-6.1 Sol was used to assist in constructing and verifying examples presented in this paper.

\appendix

\section{Proof of Theorem \ref{thm-sound}}\label{app-a}

The soundness of (R-UT), (R-$\subseteq$) and (R-$\cap$) are trivial.   

(R-SC) Let us write $\mathcal{K}_1=\hs_{\mathit{qv}(W)\setminus\mathit{qv}(Y)}$ and $\mathcal{K}_2=\hs_{\mathit{qv}(Y)\setminus\mathit{qv}(W)}$. From the assumption $\models\{X\}\ C_1\ \{W\}$, we have $\llbracket C_1\rrbracket(X)\subseteq W$. By Lemma \ref{lem-lift}, we obtain: \begin{equation}\label{eq-sc1}\left(\llbracket C_1\rrbracket\otimes I_{\mathcal{K}_2}\right)\left(X\otimes \mathcal{K}_2\right)\subseteq W\otimes \mathcal{K}_2.\end{equation}
On the other hand, it follows from the assumption $\{Y\}\ C_2\ \{Z\}$ that $\llbracket C_2\rrbracket(Y)\subseteq Z.$ Using Lemma \ref{lem-lift} again, we have: 
\begin{equation}\label{eq-sc2}\left(\llbracket C_2\rrbracket\otimes I_{\mathcal{K}_1}\right)\left(Y\otimes \mathcal{K}_1\right)\subseteq Z\otimes \mathcal{K}_1.\end{equation}
Then we obtain:
\begin{align}\label{eq-sc3}\llbracket C_1;C_2\rrbracket(X\otimes \mathcal{K}_2)&=\left(\llbracket C_2\rrbracket\otimes I_{\mathcal{K}_1}\right)\left(\left(\llbracket C_1\rrbracket\otimes I_{\mathcal{K}_2}\right)(X\otimes\mathcal{K}_2)\right)\\
&\label{eq-sc4}\subseteq \left(\llbracket C_2\rrbracket\otimes I_{\mathcal{K}_1}\right)(W\otimes\mathcal{K}_2)\\ 
&\label{eq-sc5}\subseteq \left(\llbracket C_2\rrbracket\otimes I_{\mathcal{K}_1}\right)(Y\otimes\mathcal{K}_1)\\ 
&\label{eq-sc6}\subseteq Z\otimes \mathcal{K}_1.
\end{align} Here, (\ref{eq-sc3}) is by Definition \ref{def-denot}(2), (\ref{eq-sc4}) comes from (\ref{eq-sc1}), (\ref{eq-sc5}) is by the assumption $W\otimes\mathcal{K}_2\subseteq Y\otimes K_1$, and (\ref{eq-sc6}) is from (\ref{eq-sc2}). Therefore, it holds that $$\models\left\{X\uparrow\left(\mathit{qv}(W)\cup\mathit{qv}(Y)\right)\right\}\ C_1;C_2\ \left\{Z\uparrow\left(\mathit{qv}(W)\cup\mathit{qv}(Y)\right)\right\}$$ because $X\otimes \mathcal{K}_2=X\uparrow\left(\mathit{qv}(W)\cup\mathit{qv}(Y)\right)$ and  $Z\otimes \mathcal{K}_1=Z\uparrow\left(\mathit{qv}(W)\cup\mathit{qv}(Y)\right).$

(R-$\neg$) Assume that $\models\{X\}\ C\ \{Y\}$; that is, $\llbracket C\rrbracket(X)\subseteq Y$. Then \begin{equation}\label{eq-neg}X\subseteq \llbracket C\rrbracket^{-1}(Y)\end{equation} because $\llbracket C\rrbracket$ is an unitary operator. By equation (\ref{u-neg}) in Lemma \ref{lem-qo}, we have:\begin{align}\label{eq-neg1}\llbracket C^{-1}\rrbracket(Y^\perp)&=\left(\llbracket C^{-1}\rrbracket(Y)\right)^\perp\\ &=\left(\llbracket C\rrbracket^{-1}(Y)\right)^\perp\\ \label{eq-neg3}&\subseteq X^\perp.
\end{align} Here, (\ref{eq-neg1}) comes from the unitarity of $\llbracket C^{-1}\rrbracket$, and (\ref{eq-neg3}) is obtained using (\ref{eq-neg}).

(R-$\vee$) Assume that for all $i$, $\models\{X_i\}\ C\ \{Y_i\}$. Then $\llbracket C\rrbracket(X_i)\subseteq Y_i$. By equation (\ref{u-vee}) in Lemma \ref{lem-qo} we obtain:
$$\llbracket C\rrbracket\left(\bigvee_i X_i\right)=\bigvee_i\llbracket C\rrbracket(X_i)\subseteq\bigvee_iY_i.$$ Therefore, it holds that $\models\left\{\bigvee_i X_i\right\}\ C\ \left\{\bigvee_iY_i\right\}.$

Similarly, we can prove the soundness of (Q-$\cap$).

\section{Proof of Theorem \ref{thm-sound-atom}}\label{app-b}

Assume that $\{X\}\ C\ \{Y\}$ has a proof of the form (\ref{eq-outline}). We first prove that \begin{equation}\label{eq-chain1}\models\{X_1\uparrow\mathit{qv}(X)\}\ C\ \{Y_n\uparrow\mathit{qv}(Y)\}\end{equation} by induction on $n$. The basis case of $n=1$ is trivial. Let $C^\prime=C_1;\cdots;C_{n-1}$. Then $C=C^\prime;C_n$. We assume that the conclusion is true for $C^\prime$; that is, \begin{equation}\label{eq-chain2}\models\{X_1\uparrow\mathit{qv}(X)\}\ C^\prime\ \{Y_{n-1}\uparrow\mathit{qv}(Y)\}.\end{equation} Then by the fact that (\ref{eq-outline}) is a proof, we have:\begin{align}&\label{eq-chain3}\models\{X_n\}\ C_n\ \{Y_n\},\\ \label{eq-chain4}& Y_{n-1}\uparrow (\mathit{qv}(Y_{n-1})\cup\mathit{qv}(X_n))\subseteq X_n\uparrow (\mathit{qv}(Y_{n-1})\cup\mathit{qv}(X_n)). 
\end{align} Then it follows from (\ref{eq-chain4}) that \begin{align*}Y_{n-1}\uparrow\mathit{qv}(Y)&=(Y_{n-1}\uparrow (\mathit{qv}(Y_{n-1})\cup\mathit{qv}(X_n)))\uparrow\mathit{qv}(Y)\\ 
&\subseteq (X_n\uparrow (\mathit{qv}(Y_{n-1})\cup\mathit{qv}(X_n)))\uparrow\mathit{qv}(Y)\\ &=X_n\uparrow\mathit{qv}(Y).
\end{align*} Thus, we can apply the proof rule (R-SC) to (\ref{eq-chain2}) and (\ref{eq-chain3}) and obtain (\ref{eq-chain1}). Therefore, (\ref{eq-chain1}) is proved for all $n$. 
Finally, by the condition (1) in Definition \ref{def-outline} we assert that $\models \{X\}\ C\ \{Y\}$. 

\section{Proof of Theorem \ref{thm-sound-parallel}}\label{app-c}

Assume that the Hoare triples $\{X_i\}\ C_i\ \{Y_i\}$ $(i=1,...,n)$ have interference free proofs. We are going to prove the correctness of parallel composition $C_1\|\cdots\|C_n$: $$\models\left\{\bigwedge_{i=1}^n X_i\right\}\ C_1\|\cdots\|C_n\ \left\{\bigwedge_{i=1}^nY_i\right\}$$ that is, for any input state $|\psi\rangle\in\bigwedge_{i=1}^nX_i$, we have: $$\llbracket C_1\|\cdots\|C_n\rrbracket|\psi\rangle\subseteq\bigwedge_{i=1}^nY_i.$$ By Definition \ref{def-p-denot}, it suffices to show that whenever \begin{equation}\label{eq-proof-p1}(C_1\|\cdots\|C_n,|\psi\rangle)\rightarrow^\ast(\downarrow,|\varphi\rangle)\end{equation} then the final state $|\varphi\rangle\in\bigwedge_{i=1}^nY_i$, where $\rightarrow^\ast$ stands for the reflexive and transitive closure of transition relation $\rightarrow$.   
To this end, we first prove the following:\begin{itemize}\item \textbf{Claim}: Let $Q=\bigcup_{i=1}^n\mathit{qv}(X_i)$. If \begin{equation}\label{eq-proof-p2}(C_1\|\cdots\|C_n,|\psi\rangle)\rightarrow^\ast(D_1\|\cdots\|D_n,|\varphi\rangle)\end{equation} then:\begin{enumerate}\item whenever $D_i=C_i[j)$, we have $|\psi\rangle\in\mathit{pre}(C[j))\uparrow Q$; \item whenever $D_i=\ \downarrow$, we have $|\varphi\rangle\in Y_i\uparrow Q.$
\end{enumerate} 
\end{itemize} 

We proceed by induction on the length $k$ of the transition sequence (\ref{eq-proof-p2}). For the basis case of $k=0$, we have $D_i=C_i=C[1)$, $\mathit{pre}(C_i[1))=X_i$ and $|\varphi\rangle=|\psi\rangle$. It follows from the assumption $$|\psi\rangle\in\bigwedge_{r=1}^nX_r=\bigcap_{r=1}^n(X_r\uparrow Q)$$ that $$|\varphi\rangle\in X_i\uparrow Q=\mathit{pre}(C_i[1))\uparrow Q.$$ Thus, the conclusion is true for $k=0$. 

Now we assume that the conclusion is true for the case of $<k$ as the induction hypothesis. Let us further consider the case of $k$. First, transition sequence (\ref{eq-proof-p2}) can be rewritten as 
\begin{equation}\label{eq-proof-p3}(C_1\|\cdots\|C_n,|\psi\rangle)\rightarrow^{k-1}(E_1\|\cdots\|E_n,|\eta\rangle)\rightarrow(D_1\|\cdots\|D_n,|\varphi\rangle)\end{equation} where $\rightarrow^{k-1}$ denotes the composition of $k-1$ copies of transition relation $\rightarrow$. By the transition rule (PC) for parallel composition, we can assume that the last step of (\ref{eq-proof-p3}) is derived as follows:
\begin{equation}\label{eq-middle}\frac{(E_l,|\eta\rangle)\rightarrow (D_l,|\varphi\rangle)}{(E_1\|\cdots\|E_n,|\eta\rangle)\rightarrow (E_1\|\cdots\|E_{l-1}\|D_l\|E_{l+1}\|\cdots\|E_n,|\varphi\rangle)}
\end{equation} where $E_l=C_{lr};D_l$ and $C_{lr}$ is a basic quantum gate or an atomic region. Moreover, it follows from the transition $(E_l,|\eta\rangle)\rightarrow (D_l,|\varphi\rangle)$ in the premise of (\ref{eq-middle}) that 
\begin{equation}\label{eq-proof-p4}(C_{lr},|\eta\rangle)\rightarrow (\downarrow, |\varphi\rangle).
\end{equation} Now we consider the following cases:

\textbf{Case 1}: For $i=l$, if $D_i=C_i[j)$, then from $E_l=C_{lr};D_l$ we obtain that $r=j-1$ and $$E_i=C_{i(j-1)};C_i[j)=C_i[j-1).$$ Then by the induction hypothesis, it holds that 
\begin{equation}\label{eq-proof-p5}|\eta\rangle\in\mathit{pre}(C_i[j-1))\uparrow Q=X_{i(j-1)}\uparrow Q.\end{equation} On the other hand, in the proof of $C_i$, we have:
\begin{equation}\label{eq-proof-p6}\left\{X_{i(j-1)}\right\}\ C_{i(j-1)}\ \left\{Y_{i(j-1)}\right\}.\end{equation} Using Lemma \ref{lem-clift} we can derive from (\ref{eq-proof-p4}), (\ref{eq-proof-p5}) and (\ref{eq-proof-p6}) that \begin{align}
|\varphi\rangle\in Y_{i(j-1)}\uparrow Q&=\left(Y_{i(j-1)}\uparrow\left(\mathit{qv}(Y_{i(j-1)})\cup\mathit{qv}(X_{ij})\right)\right)\uparrow Q\\ 
\label{eq-proof-p7}&\subseteq\left(X_{ij}\uparrow\left(\mathit{qv}(Y_{i(j-1)})\cup\mathit{qv}(X_{ij})\right)\right)\uparrow Q\\ &=X_{ij}\uparrow Q=\mathit{pre}(C_i[j))\uparrow Q.
\end{align} Here, (\ref{eq-proof-p7}) comes from the condition (3) in Definition \ref{def-outline}. Similarly, using the condition (1) in Definition \ref{def-outline} we can show that if $D_i=\ \downarrow$, then $|\varphi\rangle\in Y_i\uparrow Q$.  

\textbf{Case 2}: For $i\neq l$, if $D_i=C_i[j)$, then we have $E_i=D_i=C_i[j)$ from (\ref{eq-proof-p3}). By the induction hypothesis, it holds that \begin{equation}\label{eq-inter1}|\eta\rangle\in\mathit{pre}(C_i[j))\uparrow Q=X_{ij}\uparrow Q.
\end{equation} On the other hand, since $E_l=C_{lr};D_l$, we have $D_l=C_l[r+1)$ and $E_l=C_l[r)$. By the induction hypothesis again, we obtain:\begin{equation}\label{eq-inter2}|\eta\rangle\in\mathit{pre}(C_l[r))\uparrow Q=X_{lr}\uparrow Q.\end{equation}
Combining (\ref{eq-inter1}) and (\ref{eq-inter2}), we have:\begin{align}\label{eq-inter12}|\eta\rangle\in(X_{ij}\uparrow Q)\cap(X_{lr}\uparrow Q)=(X_{ij}\wedge X_{lr})\uparrow Q. 
\end{align} The equality in (\ref{eq-inter12}) can be easily derived using Lemma \ref{lem-cap}. Now the interference freedom assumption of the proofs implies:
\begin{equation}\models\{X_{ij}\wedge X_{lr}\}\ C_{lr}\ \{X_{ij}\}.\end{equation} By Lemma \ref{lem-clift}, we assert: 
\begin{equation}\label{eq-inter3}\models\{(X_{ij}\wedge X_{lr})\uparrow Q\}\ C_{lr}\ \{X_{ij}\uparrow Q\}.\end{equation}
Combining from (\ref{eq-proof-p3}), (\ref{eq-inter12}) and (\ref{eq-inter3}) leads to $$|\varphi\rangle\in X_{ij}\uparrow Q=\mathit{pre}(C_i[j))\uparrow Q.$$ Similarly, we can prove that if $D_i=\ \downarrow$, then $|\varphi\rangle\in Y_i\uparrow Q.$

Finally, by the \textbf{Claim} (2), we see that the state $|\varphi\rangle$ in (\ref{eq-proof-p1}) satisfies: $$|\varphi\rangle\in\bigcap_{i=1}^n (Y_i\uparrow Q)=\bigwedge_{i=1}^nY_i$$ and the proof is completed. 

\end{document}